\documentclass[preprintnumbers,amsmath,amssymbm,prd]{revtex4}
\usepackage{epsfig}
\usepackage{graphicx}
\usepackage{amssymb}

\begin{document}

\title{A Compact theorem on the compactness of ultra-compact objects with monotonically decreasing matter fields}
\author{Shahar Hod}
\address{The Ruppin Academic Center, Emeq Hefer 40250, Israel}
\address{}
\address{The Hadassah Institute, Jerusalem 91010, Israel}
\date{\today}

\begin{abstract}
\ \ \ Self-gravitating horizonless ultra-compact objects that possess light rings 
have attracted the attention of physicists and mathematicians in recent years. 
In the present compact paper we raise the following physically interesting question: 
Is there a lower bound on the global compactness parameters ${\cal C}\equiv\text{max}_r\{2m(r)/r\}$ of spherically 
symmetric ultra-compact objects? 
Using the non-linearly coupled Einstein-matter field equations we explicitly prove that spatially regular 
ultra-compact objects with monotonically decreasing density functions (or monotonically decreasing 
radial pressure functions) are characterized by the lower bound ${\cal C}\geq1/3$ on their dimensionless compactness 
parameters. 
\end{abstract}
\bigskip
\maketitle

\section{Introduction}

One of the most important predictions of the Einstein field equations is the existence of light rings in curved spacetimes of 
compact astrophysical objects \cite{Bar,Chan,ShTe,Notebhlr,Lu1,Hodrec,YY}. 
In particular, as explicitly proved in \cite{Hodlb}, spherically symmetric 
black-hole spacetimes have at least one closed null circular geodesic whose radius $r_{\gamma}$ is larger 
than the horizon radius $r_{\text{H}}$ of the central black hole. 
A generalization of this theorem to non-spherically symmetric black-hole spacetimes 
has been presented in the physically important work \cite{Herne}. 

Intriguingly, it is well established in the physics literature (see 
\cite{Hodt0,Kei,Mag,Hodmz1,CBH,Hodmz2,Zde1,Hodlbb} and references therein) 
that horizonless ultra-compact objects, which describe spatially regular configurations of self-gravitating matter fields, 
may also possess closed light rings. 
The dimensionless compactness parameters \cite{Noteunits}
\begin{equation}\label{Eq1}
{\cal C}\equiv\text{max}_r\Big\{{{2m(r)}\over{r}}\Big\}\
\end{equation}
of horizonless ultra-compact objects are characterized by the inequality ${\cal C}<1$ 
[here $m(r)$ is the mass of the self-gravitating matter fields which is contained within a sphere of radius $r$]. 
The relation (\ref{Eq1}) implies that horizonless ultra-compact objects are less compact than black holes \cite{Notechh}. 

Interestingly, it has been proved that in some situations the compactness 
parameters of ultra-compact objects can be bounded from below. 
In particular, it has been explicitly proved in \cite{Hodt0} that the compactness parameters of 
horizonless self-gravitating objects whose 
matter fields are characterized by a non-negative
energy-momentum trace ($T\geq0$ \cite{Bond1,Notetr}) are bounded from below by the relation 
\begin{equation}\label{Eq2}
{\cal C}\geq{2\over3}\ \ \ \ \text{for ultra-compact objects with $T\geq0$}\  .
\end{equation}
In addition, the compactness parameters of isotropic ultra-compact objects that possess closed light rings 
are bounded from below by the relation \cite{Hodlbb,Noteat,Notest}
\begin{equation}\label{Eq3}
{\cal C}\geq{{1}\over{2}}\ \ \ \ \text{for isotropic ultra-compact objects}\  .
\end{equation}

Following \cite{Nrp} (see also references therein), in the present compact paper we shall consider generic 
(that is, not necessarily isotropic) self-gravitating 
matter configurations which are characterized by monotonically decreasing 
energy densities or monotonically decreasing radial pressures. 
In particular, using {\it analytical} techniques we shall reveal the existence of a previously unknown 
lower bound on the dimensionless compactness parameters of these horizonless 
ultra-compact objects \cite{Notett}.

\section{Description of the system}

We consider horizonless curved spacetimes of spatially regular ultra-compact objects that 
possess null circular geodesics. 
The line element that characterizes the spherically symmetric spacetime can be expressed, 
using the Schwarzschild spacetime coordinates $\{t,r,\theta,\phi\}$, in the form 
\cite{Bekreg,Hodfast,Hodm}
\begin{equation}\label{Eq4}
ds^2=-e^{-2\delta}\mu dt^2 +\mu^{-1}dr^2+r^2(d\theta^2 +\sin^2\theta d\phi^2)\  ,
\end{equation}
where $\mu=\mu(r)$ and $\delta=\delta(r)$. 

The dimensionless metric function $\mu(r)$ of spatially regular asymptotically flat spacetimes is 
characterized by the boundary conditions \cite{Bekreg}
\begin{equation}\label{Eq5}
\mu(r\to0)\to1\
\end{equation}
and
\begin{equation}\label{Eq6}
\mu(r\to\infty)\to1\  .
\end{equation}
For spatially regular asymptotically flat spacetimes the metric function $\delta(r)$ is 
characterized by the boundary conditions \cite{Bekreg}
\begin{equation}\label{Eq7}
\delta(r\to0)<\infty\
\end{equation}
and
\begin{equation}\label{Eq8}
\delta(r\to\infty)\to 0\  .
\end{equation}

The Einstein-matter field equations $G^{\mu}_{\nu}=8\pi T^{\mu}_{\nu}$ of the spherically symmetric 
curved spacetime (\ref{Eq4}) can 
be expressed in the form of two non-linearly coupled differential equations \cite{Bekreg,Hodfast,Hodm},
\begin{equation}\label{Eq9}
{{d\mu}\over{dr}}=-8\pi r\rho+{{1-\mu}\over{r}}\
\end{equation}
and
\begin{equation}\label{Eq10}
{{d\delta}\over{dr}}=-{{4\pi r(\rho +p)}\over{\mu}}\  ,
\end{equation}
where the radially dependent matter parameters $\{\rho(r),p(r),p_{\text{T}}(r)\}$ \cite{Bond1}
\begin{equation}\label{Eq11}
\rho\equiv-T^{t}_{t}\ \ \ \ ,\ \ \ \ p\equiv T^{r}_{r}\ \ \ \ , \ \ \ \ p_T\equiv T^{\theta}_{\theta}=T^{\phi}_{\phi}\
\end{equation}
are respectively the energy density, the radial pressure, and the tangential pressure \cite{Notept} of 
the self-gravitating fields. 
We shall assume that the matter fields respect the dominant energy condition which implies the pressure-density 
relation \cite{Bekreg}
\begin{equation}\label{Eq12}
0\leq|p|,|p_{\text{T}}|\leq\rho\  .
\end{equation}

\section{Lower bound on the compactness parameters of ultra-compact objects with monotonically 
decreasing energy densities (or monotonically decreasing radial pressures)}

In the present section we shall explicitly prove, using the Einstein-matter field equations, 
that the dimensionless compactness parameters [see Eq. (\ref{Eq1})] of 
spatially regular horizonless ultra-compact objects 
whose radially dependent density functions (or radially dependent pressure functions) are 
monotonically decreasing are bounded from below. 
(See the excellent review \cite{Nrp} and references therein for interesting star 
models that are characterized by this monotonicity property). 
 
We first note that, as explicitly proved in \cite{Hodhair}, the null circular geodesics of the 
spherically symmetric curved spacetime (\ref{Eq4}) are characterized by the functional relation
\begin{equation}\label{Eq13}
{\cal N}(r=r_{\gamma})=0\  ,
\end{equation}
where the dimensionless function ${\cal N}(r)$ is given by the pressure dependent expression 
\begin{equation}\label{Eq14}
{\cal N}(r)\equiv 3\mu-1-8\pi r^2p\  .
\end{equation}

Taking cognizance of the Einstein equation (\ref{Eq9}) one obtains the functional relation 
\begin{equation}\label{Eq15}
\mu(r)=1-{{2m(r)}\over{r}}\
\end{equation}
for the dimensionless metric function, where the gravitational mass $m(r)$ 
which is contained within a sphere of radius $r$ is given by the integral relation 
\begin{equation}\label{Eq16}
m(r)=\int_{0}^{r} 4\pi x^{2}\rho(x)dx\  .
\end{equation}

For ultra-compact objects whose density functions are monotonically decreasing,  
\begin{equation}\label{Eq17}
{{d\rho}\over{dr}}\leq0\  ,
\end{equation}
one can use the integral relation (\ref{Eq16}) in order to derive the lower bound 
\begin{equation}\label{Eq18}
m(r_{\gamma})=\int_{0}^{r_{\gamma}} 4\pi x^{2}\rho(x)dx\geq
\rho(r_{\gamma})\cdot\int_{0}^{r_{\gamma}} 4\pi x^{2}dx=
\rho(r_{\gamma})\cdot{{4\pi}\over{3}}r^{3}_{\gamma}\
\end{equation}
on the gravitational mass which is contained within the radius of the light ring. 
Using the dominant energy condition (\ref{Eq12}) one obtains from (\ref{Eq18}) 
the relation 
\begin{equation}\label{Eq19}
m(r_{\gamma})\geq p(r_{\gamma})\cdot{{4\pi}\over{3}}r^{3}_{\gamma}\  .
\end{equation}

Likewise, for ultra-compact objects whose radial pressure functions are monotonically decreasing,  
\begin{equation}\label{Eq20}
{{dp}\over{dr}}\leq0\  ,
\end{equation}
one can use the integral relation (\ref{Eq16}) and the dominant energy condition (\ref{Eq12}) 
in order to derive the lower bound 
\begin{equation}\label{Eq21}
m(r_{\gamma})=\int_{0}^{r_{\gamma}} 4\pi x^{2}\rho(x)dx\geq
\int_{0}^{r_{\gamma}} 4\pi x^{2}p(x)dx\geq
p(r_{\gamma})\cdot\int_{0}^{r_{\gamma}} 4\pi x^{2}dx=
p(r_{\gamma})\cdot{{4\pi}\over{3}}r^{3}_{\gamma}\  .
\end{equation}

From Eqs. (\ref{Eq19}) and (\ref{Eq21}) one deduces the dimensionless inequality 
\begin{equation}\label{Eq22}
{{m(r_{\gamma})}\over{r_{\gamma}}}\geq
{{4\pi}\over{3}}p(r_{\gamma})r^{2}_{\gamma}\  .
\end{equation}
Taking cognizance of Eqs. (\ref{Eq13}), (\ref{Eq14}), (\ref{Eq15}), and (\ref{Eq22}) 
one obtains the relation
\begin{equation}\label{Eq23}
0=3\mu(r_{\gamma})-1-8\pi r^2_{\gamma}p(r_{\gamma})=2-{{6m(r_{\gamma})}\over{r_{\gamma}}}
-8\pi r^2_{\gamma}p(r_{\gamma})\leq2-16\pi r^2_{\gamma}p(r_{\gamma})\  ,
\end{equation}
which yields the characteristic inequality
\begin{equation}\label{Eq24}
8\pi r^2_{\gamma}p(r_{\gamma})\leq1\
\end{equation}
at the radial location of the light ring. 

Substituting the analytically derived relation (\ref{Eq24}) into Eqs. (\ref{Eq13}) and (\ref{Eq14}) one 
finds the dimensionless inequality
\begin{equation}\label{Eq25}
\mu(r_{\gamma})\leq{2\over3}\
\end{equation}
for the metric function, which yields the mass-radius lower bound [see Eq. (\ref{Eq15})]
\begin{equation}\label{Eq26}
{{m(r_{\gamma})}\over{r_{\gamma}}}\geq{1\over 6}\
\end{equation}
at the radial location of the light ring. 

\section{Summary and discussion}

Horizonless spacetimes of spatially regular self-gravitating matter configurations 
may share some physically interesting properties with curved black-hole spacetimes. 
In particular, despite the fact that the characteristic compactness parameters of 
horizonless ultra-compact objects are smaller 
than the corresponding compactness parameter ${\cal C}_{\text{BH}}=1$ that characterizes 
spherically symmetric black holes \cite{Notechh}, 
horizonless compact objects may possess light rings on which massless particles can perform closed circular motions (see 
\cite{Hodt0,Kei,Mag,Hodmz1,CBH,Hodmz2,Zde1,Hodlbb} and references therein). 

In the present compact paper we have quantified the meaning of the term `ultra-compact' which is used 
in the physics literature to describe self-gravitating horizonless matter configurations that possess null circular geodesics. 
In particular, we have addressed the following 
physically interesting question: How compact are self-gravitating 
ultra-compact objects with closed light rings? 

The main analytical results derived in this paper and their physical implications are as follows:

(1) Using analytical techniques we have explicitly proved that the compactness parameters of horizonless 
ultra-compact objects whose matter fields are characterized by monotonically decreasing 
energy densities or monotonically decreasing radial pressures (see \cite{Nrp} and references therein 
for physically interesting models of compact objects that are characterized by this monotonicity property) 
are bounded from below by the dimensionless relation [see Eqs. (\ref{Eq1}) and (\ref{Eq26})] 
\begin{equation}\label{Eq27}
{\cal C}\geq {{1}\over{3}}\  .
\end{equation}

(2) It is worth emphasizing the fact that, as opposed to the bound (\ref{Eq3}) whose 
validity is restricted to the case of isotropic ultra-compact objects \cite{Hodlbb}, 
the newly derived dimensionless lower bound (\ref{Eq27}) 
on the compactness parameters of ultra-compact objects 
is valid for all spatially regular horizonless matter configurations (not necessarily isotropic) with light rings 
that are characterized by monotonically decreasing energy densities or monotonically decreasing 
radial pressure functions. 

(3) It is physically interesting to note that the analytically derived lower bounds (\ref{Eq3}) and (\ref{Eq27}) 
on the compactness parameters of ultra-compact objects imply that if a neutron star with 
the typical compactness value ${\cal C}\sim0.4$ \cite{CP} has light rings then it must be characterized by 
non-isotropic internal pressures. 

\bigskip
\noindent {\bf ACKNOWLEDGMENTS}

This research is supported by the Carmel Science Foundation. I thank
Yael Oren, Arbel M. Ongo, Ayelet B. Lata, and Alona B. Tea for
stimulating discussions.



\begin{thebibliography}{99}
\bibitem{Notebhlr} See \cite{Lu1,Hodrec,YY} and references therein for recent studies 
of the physical and mathematical properties of curved spacetimes that contain closed light rings.
\bibitem{Noteunits} We use natural units in which $G=c=1$.
\bibitem{Notechh} Note that spherically symmetric black holes are 
characterized by the dimensionless compactness relation
$2m(r_{\text{H}})/r_{\text{H}}=1$, where $\{r_{\text{H}},m(r_{\text{H}})\}$ are 
respectively the radius of the black-hole horizon and the gravitational mass contained within the horizon.
\bibitem{Notetr} The trace of the energy-momentum tensor is 
given by the relation $T=-\rho+p_r+2p_{\text{T}}$, where the spatially dependent functions 
$\{\rho,p_r,p_{\text{T}}\}$ are respectively the energy density, the radial pressure, and the
tangential pressure of the self-gravitating matter configurations \cite{Bond1}.
\bibitem{Noteat} Note that the bound (\ref{Eq3}) is valid for both possible signs of the 
trace of the energy-momentum tensor that characterizes the self-gravitating matter fields.
\bibitem{Notest} It is worth noting that the stronger lower bound ${\cal C}\geq7/12$ on the compactness 
parameter (\ref{Eq1}) has been derived in \cite{Hodlbb} for isotropic ultra-compact objects 
which are characterized by negative ($T<0$) energy-momentum traces.
\bibitem{Notett} It is worth emphasizing that the new lower bound, to be explicitly derived below [see 
Eq. (\ref{Eq27}) below], is valid for both possible signs of the energy-momentum trace $T$ that 
characterizes the self-gravitating matter fields.
\bibitem{Notept} Note that the tangential pressure $p_{\text{T}}$ appears in the 
Tolman–Oppenheimer–Volkoff (TOV) equation 
$dp/dr=\big[(3\mu-1-8\pi r^2p)(\rho+p)+2\mu(-\rho-3p+2p_{T})\big]/(2\mu r)$. 
It is worth emphasizing the fact that our theorem for the compactness of ultra-compact objects 
with monotonically decreasing matter fields does not use the TOV equation.
\end{thebibliography}
\end{document}